\documentclass[11pt]{article}
\usepackage[utf8]{inputenc}

\newcommand{\onbb}      {\ensuremath{0\nu\beta\beta}}

\newcommand{\ctsper}    {cts/(keV\,kg\,yr)}

\newcommand{\MJD}       {\textsc{Majorana Demonstrator}}
\newcommand{\Gerda}     {\textsc{Gerda}}

\newcommand{\be}        {\begin{equation}}
\newcommand{\ee}        {\end{equation}}

\usepackage[margin=1.in]{geometry}
\usepackage{amssymb}
\usepackage{amsmath}
\usepackage{cite}
\usepackage{bm}
\usepackage{graphicx}
\usepackage{xcolor}
\usepackage{authblk}

\title{The Front-End electronics for the liquid Argon instrumentation of the LEGEND-200 experiment}

\author[1]{I.~Abritta Costa}
\author[2]{A.~Budano}
\author[1]{N.~Burlac}
\author[2]{F.~Paissan}
\author[1]{G.~Salamanna\thanks{corresponding author}}
\author[2]{D.~Tagnani}
\affil[1]{Department of Mathematics and Physics, Roma Tre University, Rome, Italy}
\affil[2]{Sezione INFN di Roma Tre, Rome, Italy}

\begin{document}
\maketitle
\begin{abstract}
In this paper we provide a detailed technical description of the Front-End (FE) electronics for the liquid Argon instrumentation of the LEGEND-200 experiment, searching for the very rare, hypothetical neutrinoless double $\beta$ decay process at the Italian Laboratori Nazionali del Gran Sasso. The design stems from the need to read out the silicon photo-multiplier response to the scintillation light in the liquid Argon with excellent single-photon resolution. The FE electronics is required to be placed far from the detectors to meet the experiment's radio-purity constraints. This constraint represents a challenge for a high signal-to-noise ratio. We address how this could be achieved in a stable way. The system was installed in July 2021 and has been commissioned with the rest of LEGEND-200, proving we could attain a very low overall level of electrical noise of 250 $\mu$V peak-to-peak. 
\end{abstract}

\section*{Introduction}

The search for neutrinoless double beta (\onbb) decay, a yet-to-be-observed weak transition, is a topic of broad and current interest in modern physics. Its detection would unequivocally prove the existence of new lepton-number violating physics beyond the Standard Model of particle physics. It would also establish a link to the origin of neutrino mass.



Several experiments~\cite{ref:review, ref:review ge} take advantage of various technologies to search for this rare transition using different isotopes, such as: $^{76}$Ge, $^{82}$Se, $^{100}$Mo, $^{130}$Te and $^{136}$Xe~\cite{ref:review isotopes}. The latest \Gerda~results lead the \onbb~decay field, reporting the highest sensitivity on the half-life of \onbb~decay ($1.8\times10^{26}$~yr for $^{76}$Ge~isotope) and the lowest background index at the Q-value of the decay ($5.2\times10^{-4}$~\ctsper)~\cite{ref:gerda final}. 

The \Gerda~experiment, at Laboratori Nazionali del Gran Sasso (LNGS) in Italy, used 44 kg of High-Purity Germanium (HPGe) detectors, enriched in the $^{76}$Ge isotope up to $\sim$87\% and deployed bare into ultra-pure Liquid Argon (LAr).
The use of germanium diodes allows to have not only an excellent energy resolution but also a high detection efficiency, being simultaneously the source and the detector of the double $\beta$ decay. The best energy resolution of germanium detectors was achieved by \Gerda's competitor, \MJD~experiment~\cite{ref:majorana}.
The combination of these two collaborations paved the way for the next-generation experiment, the Large Enriched Germanium Experiment for Neutrinoless Double Beta Decay (LEGEND)~\cite{ref:legend}.

\begin{figure}[t]\center
\includegraphics[width=0.6\textwidth,keepaspectratio]{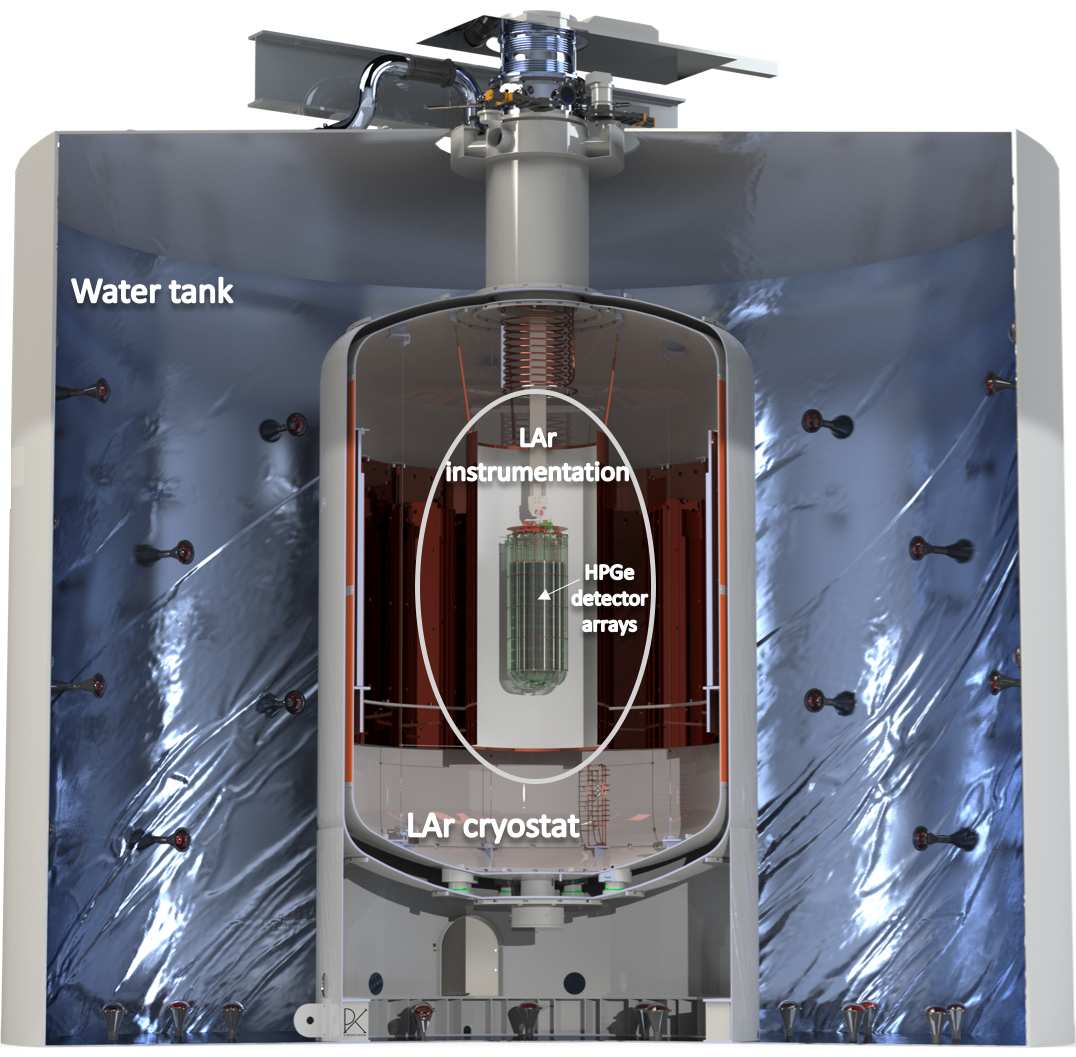}\quad\includegraphics[width=0.35\textwidth,keepaspectratio]{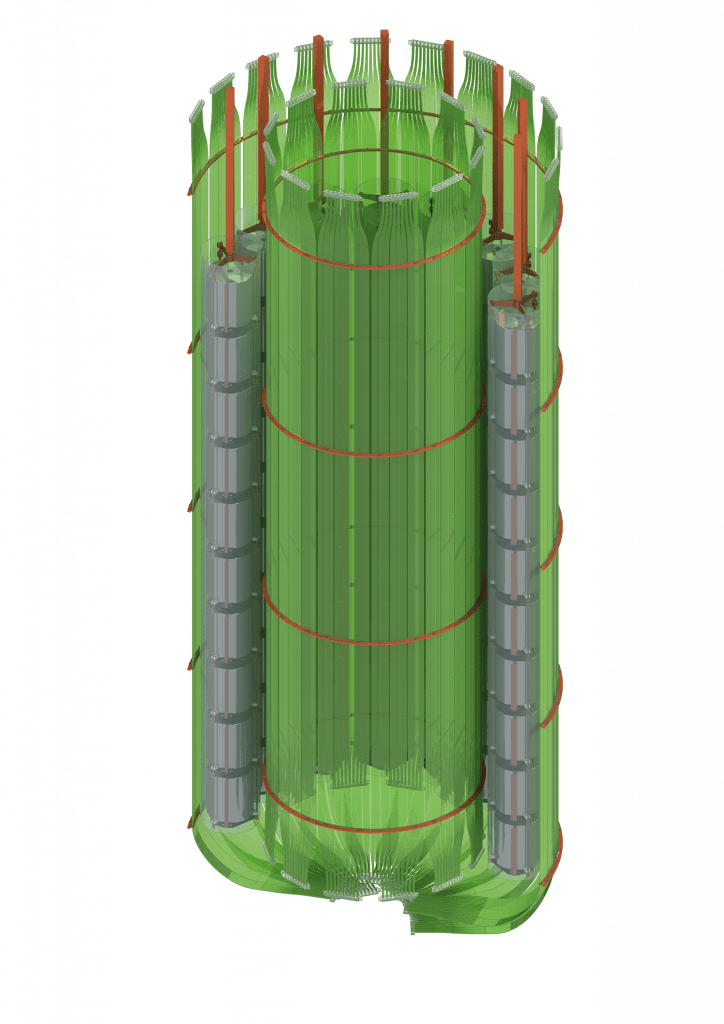}
\caption{Left: Artists view of the LEGEND-200 setup at LNGS; Right: Two concentric fiber shrouds coupled on both ends to SiPMs detectors, surround the HPGe detectors. More details in~\cite{ref:lar instr, ref:lar instr nina}.}
\label{fig:setup}
\end{figure}

\vfill

The first phase of the experiment, called LEGEND-200, takes place at LNGS. It aims to reach a sensitivity on the half-life of 10$ ^{27} $ yr by operating about 200 kg of HPGe detectors in the upgraded \Gerda~infrastructure.
To achieve such sensitivity LEGEND-200 must perform a quasi-background-free search, greatly reducing the ambient backgrounds with respect to past experiments, about a factor 3 lower than \Gerda. 

The overlying rock and the instrumented tank containing purified water remove the hadronic components of cosmic ray showers and reduce the moun flux as well as external $\gamma$ radiation. To further shield the detectors against background radiation, inside the water tank a 64~m$^3$ cryostat (a device used to maintain cryogenic temperatures) with purified LAr is present. 
The concept of an {\it active} veto, that is a region instrumented to not only absorb, but also actively detect background particles, was successfully used in \Gerda~and is replicated in LEGEND-200 \cite{ref:lar instr, ref:lar instr nina}.  For LEGEND-200 the functionalities are expanded to also measure energies and do timing, such that one should talk more appropriately of a LAr detector. The LEGEND-200 multi-layer setup is shown in Figure~\ref{fig:setup}~(Left).

\section*{The LEGEND-200 LAr detector} \label{sec:lar}

\subsection*{Light generation and detection}
The detector elements made of high-purity germanium are immersed bare in LAr (argon in the liquid phase at a temperature of 87 K). This noble liquid element is an efficient scintillator, when traversed by background particles releasing energy within. Two concentric curtains of approximately 1.5 m long Wave-Length Shifting (WLS) fibers coupled to Silicon Photo-Multiplier (SiPM) devices are able to detect scintillation light~\cite{ref:neutrino post}. The system, shown in Figure~\ref{fig:setup}~(Right), was designed to be sensitive to $\gamma$ rays from radioactive decay chains in the environment and in the detector materials; and $\alpha$ and $\beta$ decays near to or on the detector surface.

WLS is obtaining by coating the Saint Gobain BCF 91A fibers with Tetra-Phenyl Butadiene (TPB). The latter shifts the LAr scintillation light ($\lambda=$128 nm) from the vacuum ultraviolet (VUV) to blue light (440 nm) \cite{ref:tpb}, then the WLS fibers shift the light to green wavelengths. Light is read out by Ketek PM33100T SiPMs of the same type employed in GERDA Phase II. The WLS fibers are organized in modules: 20 modules for the outer barrel and 9 for the inner barrel. Each fiber module is read out by one array of SiPM at each end, in total 58 channels. 

\subsection*{Light detector arrangement and signal extraction} \label{sipm-way}
The SiPM of LEGEND-200 are arranged into groups of nine, packaged in a synthetic fused silica material which meets the radio-purity requirements. The anodes and cathodes of the nine SiPM are connected together electrically to two AXON pico-coax cables, one per each electrode. After between 20 and 190 cm (depending on the array location), the coaxial cables are connected to a 10~m long Kapton flat-band cable. In this manner, signals from the SiPM detectors are driven to the outside of the cryostat. 

\subsection*{SiPM signals to the Front-End electronics and DAQ}
From the cryostat flanges, 6~m long Cat 6 ethernet cables are run to the Front-End~(FE) electronics, where SiPM signals are read out and amplified. Each array of 9 SiPM, connected to one FE channel, produces a single waveform. The analogue pulses are then digitized by a 16-bit 62.5 MHz flash Analogue-to-Digital (ADC) converter. The data acquisition (DAQ) system of LEGEND-200 is based on the modular FlashCam system \cite{ref:fcam}, which is highly scalable and read out through Gigabit ethernet. The digital signal processing of the traces is performed within a dedicated software \footnote{https://github.com/legend-exp/pygama}.
\newline

This paper describes the middle tier of the read-out chain: the analogue FE electronics, with its receiver and amplifier stages, together with the integrated slow controls. The structure of the paper is as follows: first the physics and electronics constraints on the performance are discussed and the scheme of the electronics is described. The main body of the paper focuses on the architecture of each FE electronic board and of the FE system altogether. A dedicated paragraph offers a description of the integrated slow-control functionalities and the stability of the electrical figures provided. Finally, some figures illustrating the system performance (particularly the electronic noise level attained) are provided, which were measured after integrating the FE with the signal lines in the experiment. The overall physics performance of the LAr instrumentation is not a matter for this paper and will be discussed in a future work.


\section{FE constraints from the desired physics and the detector layout}
The design of the FE electronics for the SiPM of the LAr instrumentation derives from the performance needs and layout constraints of the detector (before the FE tier) and of the DAQ (after the FE). 
\subsection{Physics needs}
The LAr detector logic identifies background events by requesting time coincidences of signals in several SiPM arrays. A "background" event, that is an event satisfying a majority cut, is discarded from the physics data. One should maximize the veto efficiency to identify (and reject) background events and, yet, minimize the dead time and the chance to reject an interesting physics candidate event. The above principle translates into the constraint that the read-out chain be able to separate array signals originated by real charge collection at the SiPM anodes from electrical noise on the FE lines ("single-photo-electron" constraint). The goal is to design a FE electronic with a noise level low enough that a cut at a fractional charge of a single-photo-electron equivalent can be applied. This minimizes the probability of random veto coincidences and guarantees a suitably high efficiency to \onbb events. 

Another important constraint comes from the need to keep the rate of radioactive backgrounds close to the central germanium diodes of LEGEND-200 as low as possible. Materials of which detectors, read-out electronic boards and mechanical supports are made bear a level of radioactivity from the natural $^{238}$U and $^{232}$Th chains. For this reason, the SiPM FE electronics are all fit outside of the cryostat. As described in the previous sections, the overall distance between the detectors and the FE boards is of approximately 16 m. No amplification stage is allowed close to the detector elements, inside the cryostat: this adds a constraint on the maximum electrical noise that can be tolerated from the FE system.

\subsection{Mechanical constraints, differential lines}
The electrical scheme of the SiPM connections described above was preferred to a paradigm with only one coaxial cable both because of mechanical/constructive reasons and because the resulting signal read-out topology is differential. A differential scheme has the advantage of being more effective in cancelling any noise picked up along the 16 m signal propagation line. The FE is therefore designed to match the differential read-out at the receiver level. 

\section{ Electronic system for the LAr instrumentation }
\begin{figure}[ht!]\center
\includegraphics[width=0.8\textwidth,keepaspectratio]{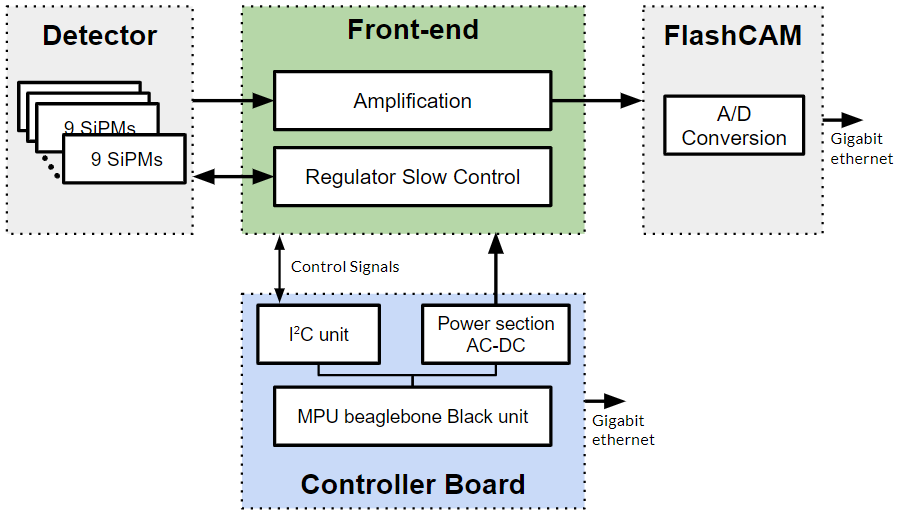}
\caption{Flowchart schematics of the full electronic system chain for the LAr instrumentation.}
\label{fig:detector2FCAM}
\end{figure}

The design of the FE chain as described in this article stems from the above physics constraints, plus the need for a smooth integration in the overall LEGEND-200 data acquisition. The developed chain consists of two sections, illustrated in Figure~\ref{fig:detector2FCAM}:
\begin{description}
    \item[Front-end:] to adapt the signal of the SiPM light detectors;
    \item[Controller Board:] for the slow control to manage the correct operation of the detector, setting the voltage bias (V$_{\text{bias}}$) and reading the values of temperature, current and voltage.
\end{description}

\begin{figure}[htb!]\center
\includegraphics[width=0.8\textwidth,keepaspectratio]{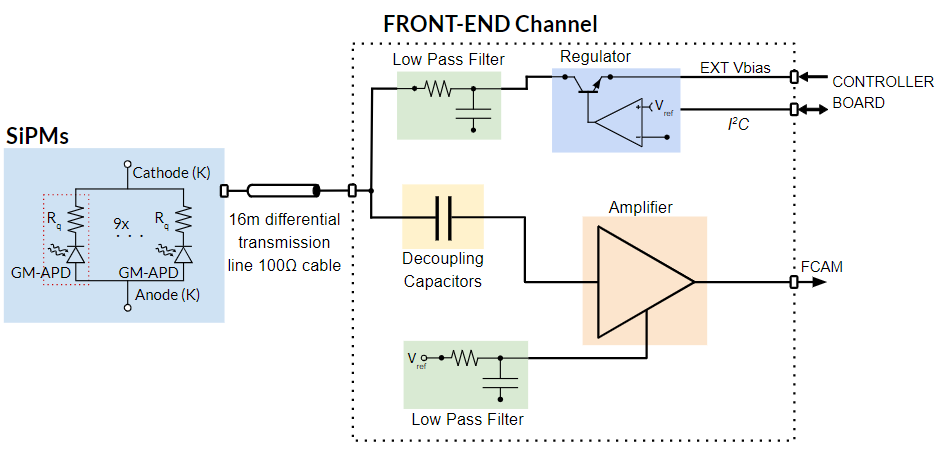}
\caption{Illustrative schematic of one channel of the Electronic front-end board.}
\label{fig:feboard}
\end{figure}

Each board accommodates 12 channels (see above for the LAr detector layout segmentation) and contains all elements necessary for the read-out and control of the SiPM module. Each board connects to power supplies according to the NIM standard, with appropriate linear adjustment sections to create a +5 V at 3 A clean power supply for the analogue section and +3.3 V at 1 A for the digital slow-control section. A primary voltage of 32 V at 20 mA and a digital communication protocol, called I$^{2}$C~\cite{ref:i2c}, are also employed, which are generated and managed by a controller board described in Sec.~\ref{controller}. An illustrative figure of a board can be found in Figure~\ref{fig:feboard}.

For the complete integration of the system with the LEGEND-200 LAr instrumentation chain, a back-plane with specific connection cables is also required. The electronic chain is designed to be acquired by the DAQ FlashCam and remotely controlled by the Central Slow-Control of the LEGEND-200 experiment.

\begin{figure}[ht!]\center
\includegraphics[width=0.7\textwidth,keepaspectratio]{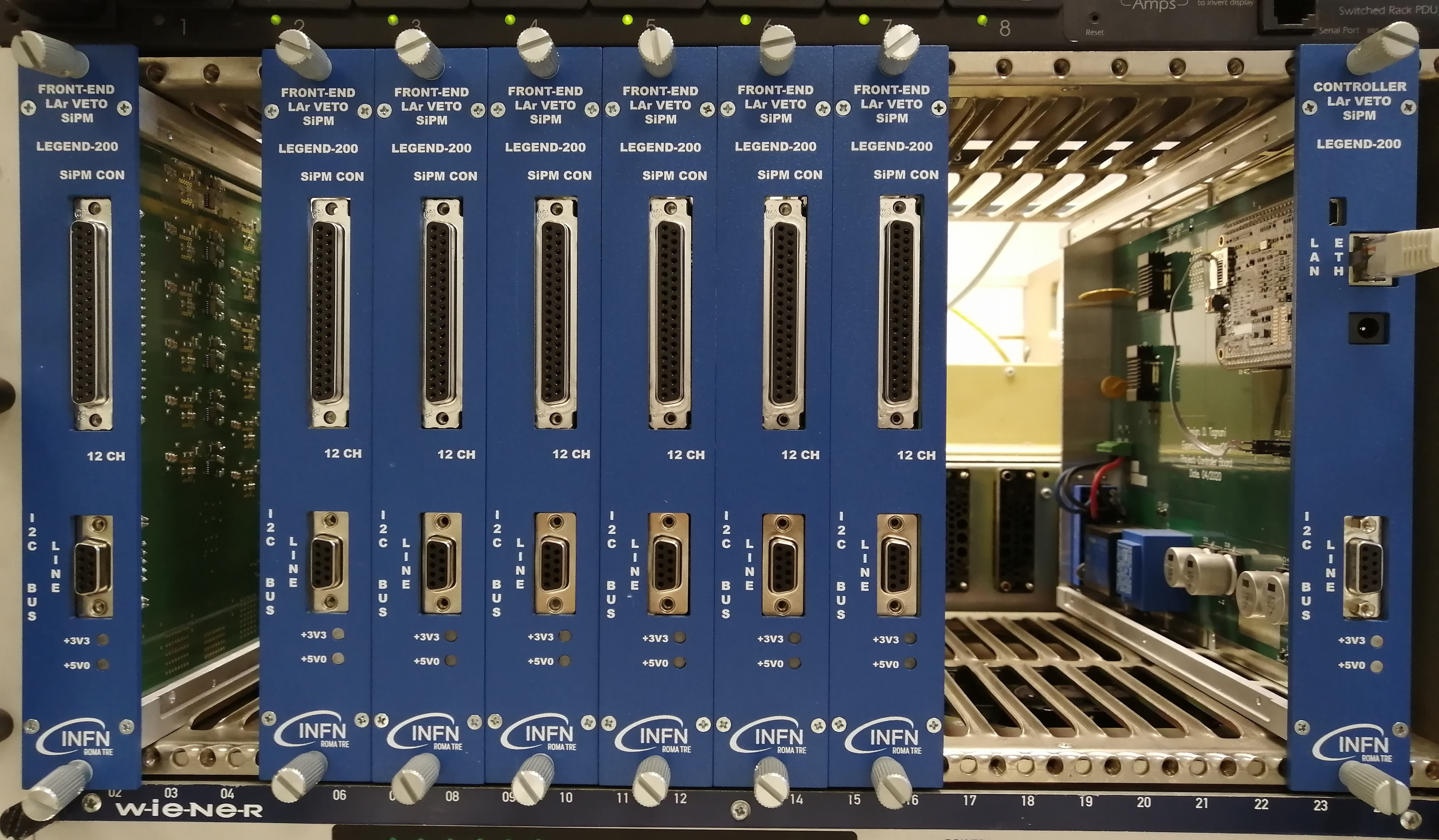}
\caption{Photo of the crate with the front-end boards and, on the right, of the controller board.}
\label{fig:crate}
\end{figure}

To allow the management of 58 channels of SiPM modules we decided to create 5 boards, plus spares, using the NIM standard. Figure~\ref{fig:crate} shows a picture of the crate with the FE boards already installed.

\subsection{Front-End Electronics boards}

\begin{figure}[htbp!]\center
\includegraphics[width=1.\textwidth,keepaspectratio]{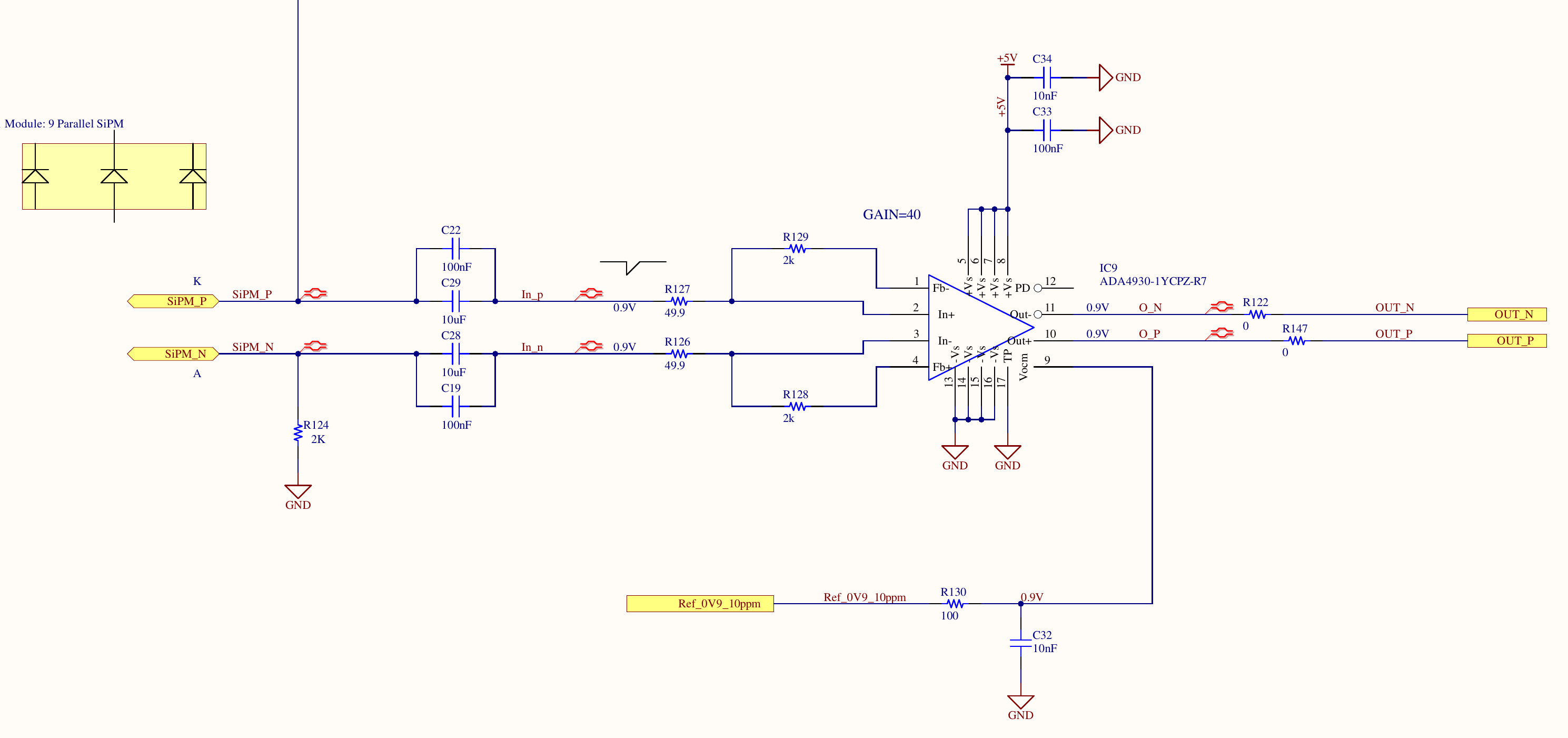}
\caption{Schematic of a differential input and output amplifier channel.}
\label{fig:SCH_Amp_DiffOut}
\end{figure}
\begin{figure}[t]\center
\includegraphics[width=0.96\textwidth,keepaspectratio]{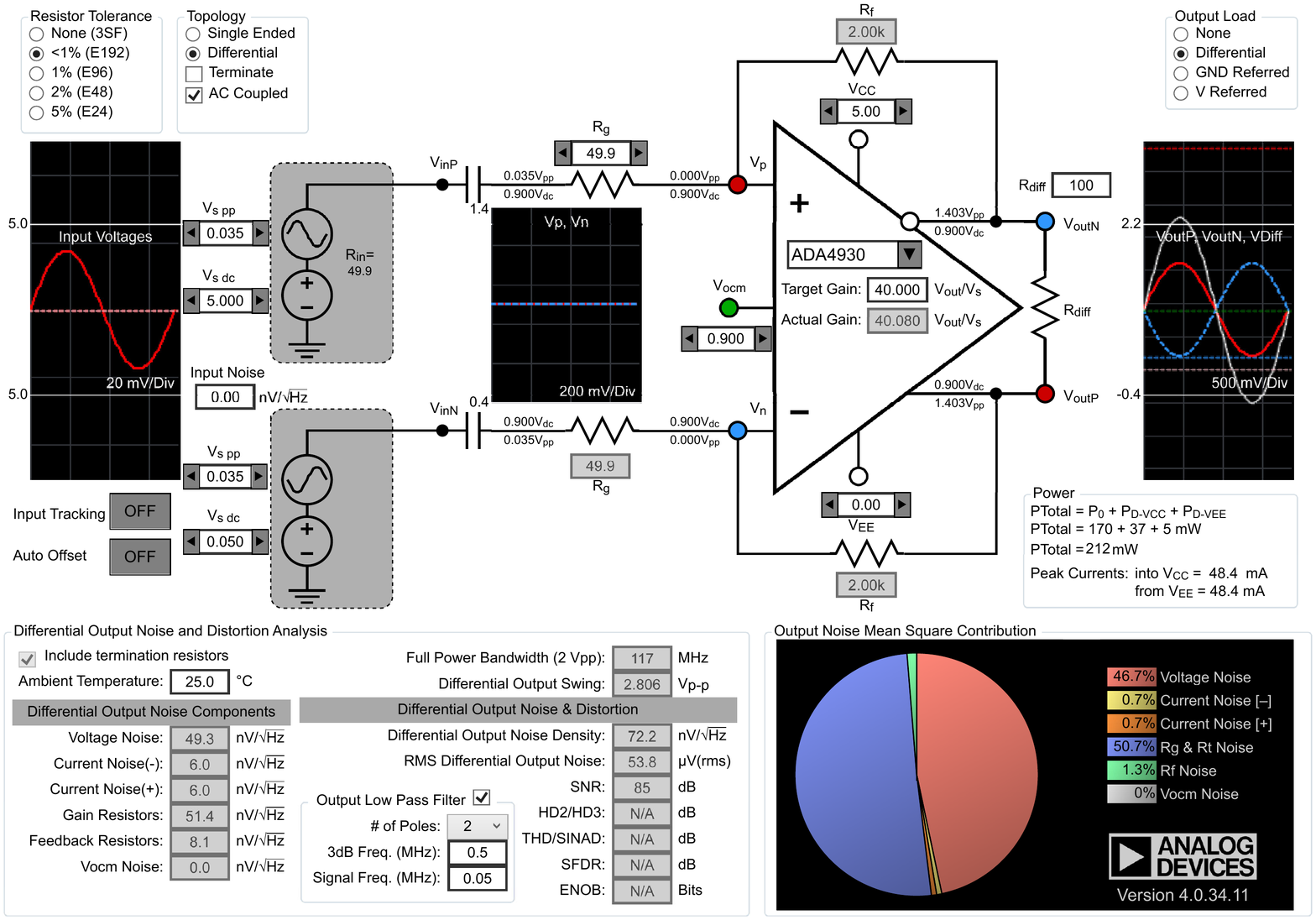} 
\caption{Simulation of the output of the FE circuit, run on the webpage https://www.analog.com/en/design-center/interactive-design-tools/adi-diffampcalc.html}
\label{fig:Simulation_Out}
\end{figure}

The schematic of a differential input and output amplifier channel is shown in Figure~\ref{fig:SCH_Amp_DiffOut}.
The C19, C22, C28 and C29 capacitors, therein referenced,  allow the AC de-coupling of the signal taken from the differential line at the regulated V$_{\text{bias}}$ voltage for each SiPM. The de-coupling capacitors are chosen with an insulation voltage of at least 100 V to avoid dielectric losses during operation. 
The amplifier chosen for the adaptation and amplification of the signal is the ADA4930-1YCPZ-R7 (element IC9 in Figure~\ref{fig:SCH_Amp_DiffOut}), with a total gain of 40. The gain on the two differential lines is defined by the ratio of the resistance of elements R129/R127 and R128/126. The ADA4930-1 is a very low noise, low distortion, high speed class of differential amplifiers. They are an ideal choice for driving 0.9 V high performance ADCs such as the one in use in the LEGEND-200 experiment. The ADA4930-1/ADA4930-2 are produced using a silicon-germanium (SiGe) technology, complementary bipolar process (see e.g. \cite{ref:comple}). They achieve a very low level of distortion with an input voltage noise of only 1.2 nV/$\sqrt{\mathrm {Hz}}$ \cite{ref:ada4930}. The chosen configuration has been simulated and studied, defining the critical points to reduce the noise as much as possible, while maintaining a band of at least 100 MHz. The result is displayed in Figure~\ref{fig:Simulation_Out} and confirms that the design is expected to meet the desired performance. In particular, a full bandwidth of 117 MHz is achieved, which allows to match in frequency the fast rise (few ns) and the slow decays (from 1-2 $\mu$s to tens of $\mu$s) of the various SiPM detectors at hand. The FE board is coupled to a second order low pass filter 
(produced by elements R130 and C32). This configuration allows to achieve a low differential output noise, with an RMS of 53.8 $\mu$V. This is satisfying for our physics needs and is compatible with what we measure in-experiment (see Sec.\ref{sec:noise-level}).

\begin{figure}[htbp!]\center
\includegraphics[width=1.\textwidth,keepaspectratio]{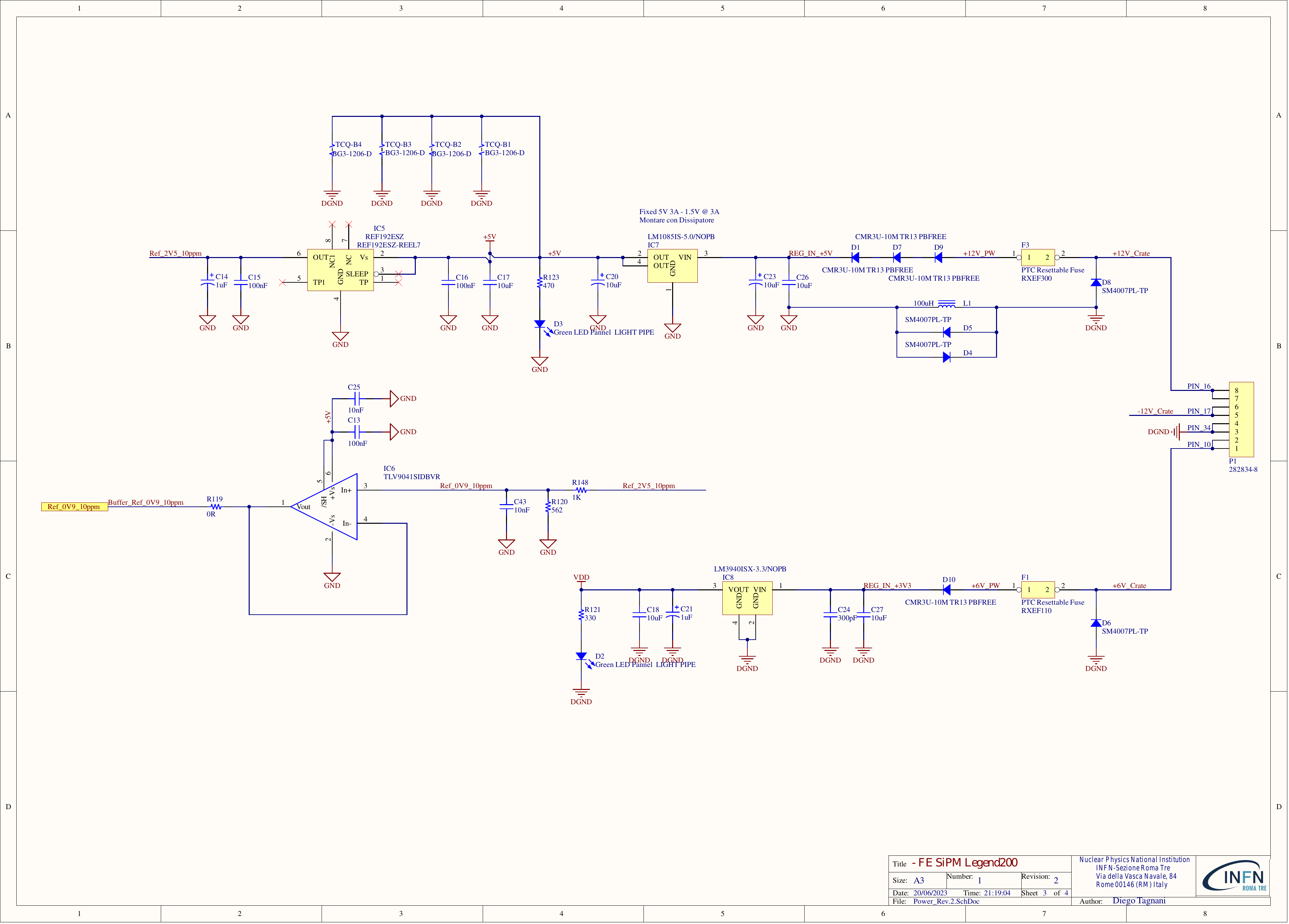}
\caption{Schematic of the power section of a front-end board.}
\label{fig:SCH_Power_FE}
\end{figure}

\begin{figure}[htbp!]\center
\includegraphics[width=1\textwidth,keepaspectratio]{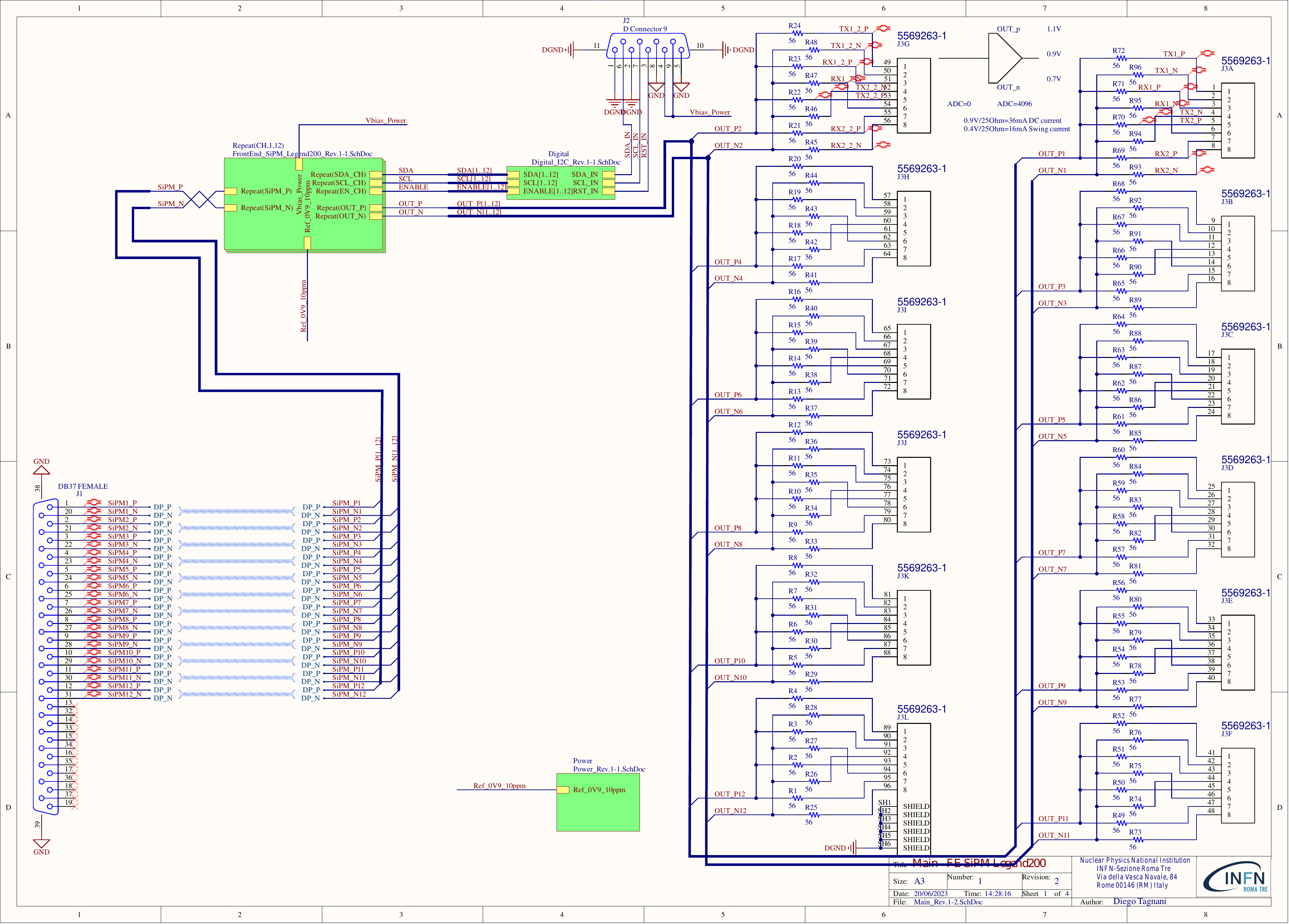}
\caption{Schematic main section of front-end board.}
\label{fig:SCH_Main_FE}
\end{figure}

The common mode voltage is set to 0.9 V, as required by the DAQ system. The common mode voltage is generated by a voltage reference integrated circuit (REF192ESZ) at 2.5 V. Its initial accuracy is $\pm{2}$ mV, while the maximum output current is 30 mA. Stability of the voltage is ensured by a linear regulator with a precision of 4 ppm/V and a temperature coefficient of 5 ppm/°C maximum. The voltage reference is followed by a passive attenuator made with resistors. Their precision is at the level of 0.1$\%$ and their temperature coefficient is of 10 ppm/°C maximum. An operational amplifier TLV9041SIDBVR in buffer configuration allows parallel distribution to the 12 channels of each board. The schematic of the power section of a front-end board is shown in Figure~\ref{fig:SCH_Power_FE}. The amplifier output has 4 differential channels in parallel on RJ45 connectors, as per FlashCam specifications. It is shown in Figure~\ref{fig:SCH_Main_FE}.

For the correct operation of a SiPM photo-detector it is necessary to provide a suitable diode voltage, stable and precise. Optimal V$_{\text{bias}}$ values for the Ketek PM33100T SiPMs of LEGEND-200 have been determined at a characterization bench at the Technical University of Munich (TUM) and range from 24 to 26 V. 
For the detector safety it is also necessary to limit the current such that it is not damaged in case of high light or mis-handling. For this reason, we have designed a power supply able to generate a V$_{\text{bias}}$ adjustable between 15 and 31.5 V, in steps of 0.05 V.

\begin{figure}[htbp!]\center
\includegraphics[width=1\textwidth,keepaspectratio]{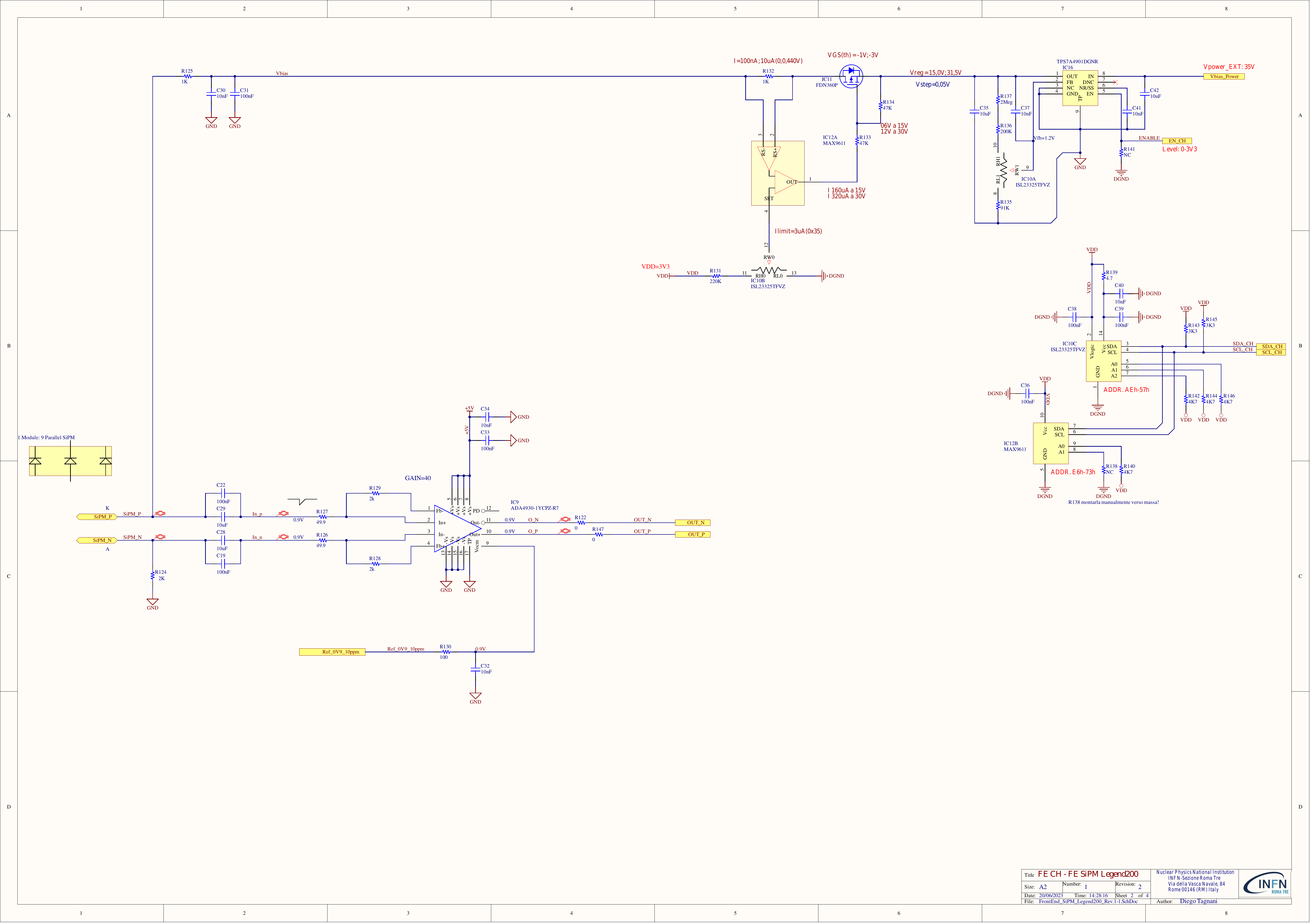}
\caption{Schematic single channel amplifier and regulation section of front-end board.}
\label{fig:SCH_Single_CH_FE}
\end{figure}

\begin{figure}[htbp!]\center
\includegraphics[width=\textwidth,keepaspectratio]{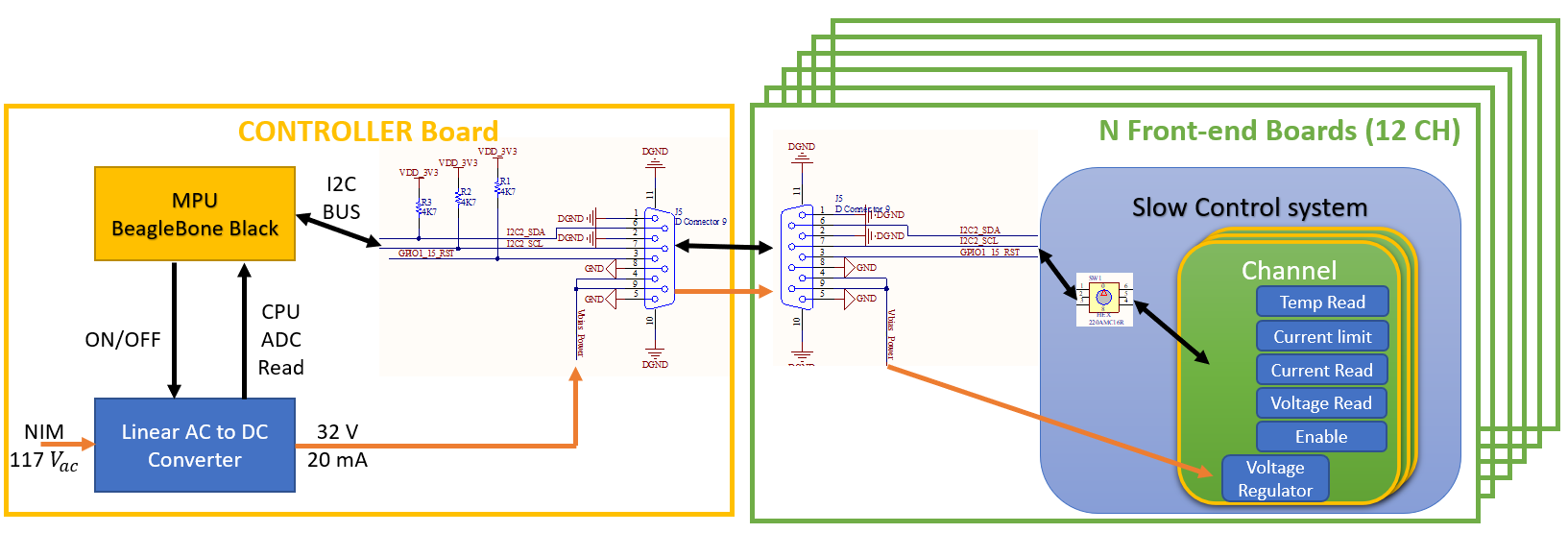}
\caption{Block diagram of front-end and controller system.}
\label{fig:BL_CH_FE}
\end{figure}

\newpage
The power supply line is filtered of any noise through the network elements R132, C31, C31 and R125. The schematic for the single channel amplifier and regulation section of a front-end board is shown in Figure~\ref{fig:SCH_Single_CH_FE}. The V$_{bias}$ generation system has several blocks, located on different boards, as shown in the block diagram in Figure~\ref{fig:BL_CH_FE}.

\begin{figure}[t]\center
\includegraphics[width=1\textwidth,keepaspectratio]{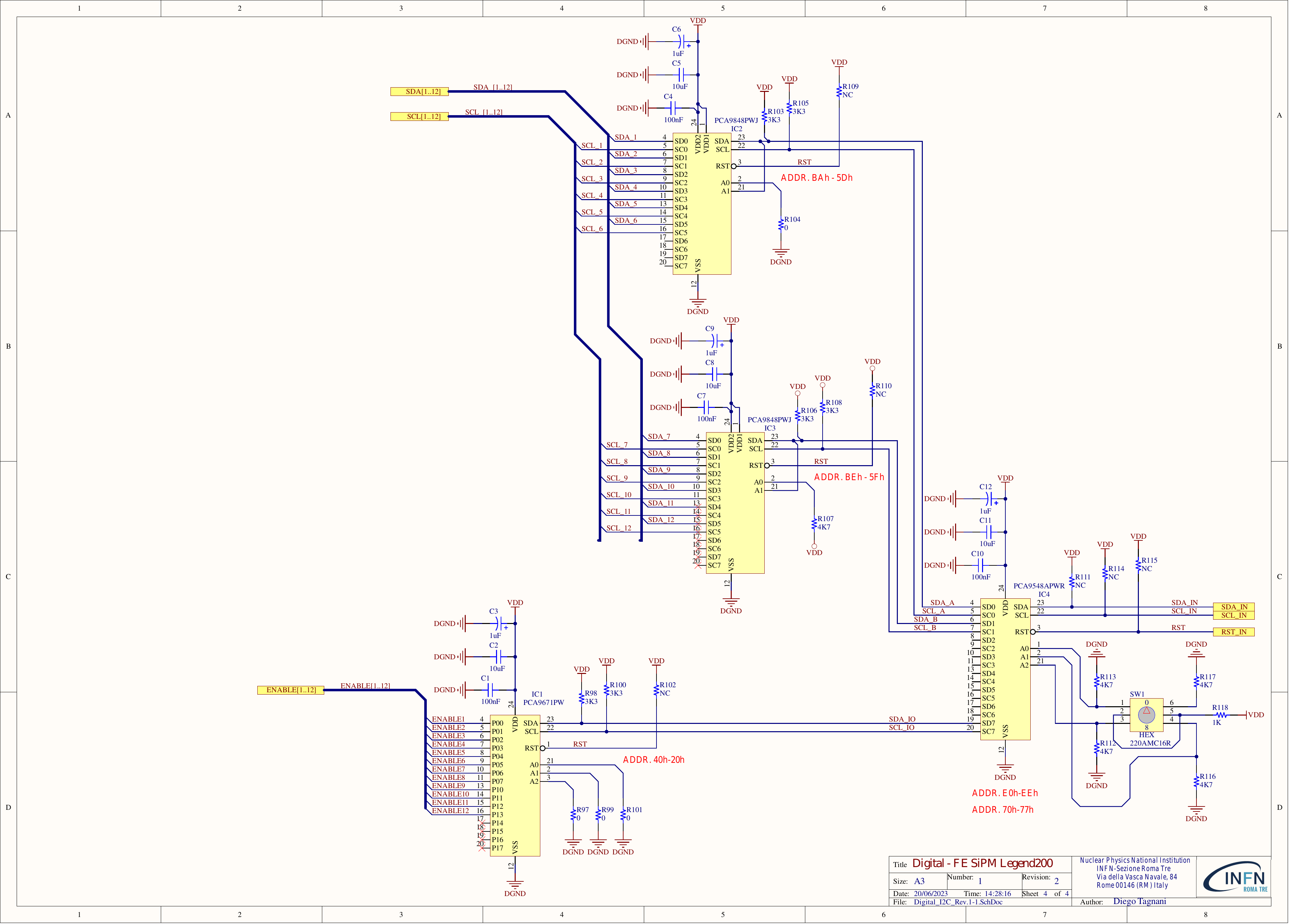}
\caption{Schematic addressing of a front-end board via I$^{2}$C bus.}
\label{fig:SCH_I2C Address_FE}
\end{figure}

\newpage
The controller board contains the processing core, a Beaglebone Black Microfluidic Processing Unit (MPU) \cite{ref:bbmpu}, which manages the entire slow control chain for all the acquisition channels, and the primary voltage generation section. Control of the FE boards is possible using components that communicate with each other through the bidirectional I$^{2}$C protocol. We designed the entire chain with two components per channel, that is 24 devices per board, plus another single-channel enable management device, for a total of 25 devices per board that need to communicate on the same digital bus. A dedicated address is assigned to each card in a unique way, through a hexadecimal rotary switch, which accommodates up to 7 cards. This allows the MPU to communicate with one board at a time, replicating the I$^{2}$C buses across the FE boards. Inside each card several elements are hosted: a PCA9548APWR device, an 8-Channel I$^{2}$C switch with a 2-bit address and two PCA9848PWJ devices, an 8-Channel I$^{2}$C switch with three bit address. These objects in cascade allow to establish a unique I$^{2}$C connection between the MPU and the specific channel control device. A PCA9671PW digital multiplexer has also been inserted in the same chain to manage the individual power on/off of the linear voltage regulator for each channel. The addressing flow of the front-end boards via the I$^{2}$C bus is shown in Figure~\ref{fig:SCH_I2C Address_FE}.

The multiplexer directly controls the enable input of the Linear Regulator used to adjust the bias voltage of each TPS7A4901DGNR channel. TPS7A49 series devices are positive, high voltage (36 V), ultra-low noise (15.4 $\mu$V RMS, 72 dB PSRR) linear regulators that can source a 150 mA load. The TPS7A49 family is designed using a bipolar technology. It is ideal for high-accuracy and high-precision applications where clean voltage rails are critical to maximize the system performance. In addition, the TPS7A49 family of linear regulators is suitable for post AC-DC converter regulation.  This device allows a regulation of the output voltage by modifying the partition network. It consists of R135, RW1, R136 and R137, that define the feedback voltage of the regulator, with V$_{FB}$ = 1.185 V. Since we want a remote regulation of the bias voltage of the SiPM, we make use of a variable resistance RW1 integrated in ISL23325TFVZ. The resistors R135, R136 and R137 define the extremes of the adjustable range, from 15 to 31.5 V. 

The I$^{2}$C Bus interface integrates two digitally controlled potentiometer (DCP) cores, wiper switches and control logic on a monolithic CMOS integrated circuit. The DCP can be used as a three-terminal potentiometer. The value of the resistance WR0 and WR1 change from 0 to 100 k$\Omega$. This voltage is regulated and filtered to reduce the noise on the supply line of the SiPM detector. It is further treated to allow the measurement of the data acquisition characteristics and to protect the same from excessive bias current that could damage it irreparably in case of exposure to the light. For this we decided to use a MAX9611AUB+T integrated circuit, which allows to monitor the parameters of the power line through the R132 sense resistance, specially chosen to be 1k$\Omega$. The MAX9611 are high-side current-sense amplifiers with an integrated 12-bit ADC and a gain block that can be configured either as an operational amplifier (op amp) or as a comparator. The high-side current-sense amplifiers operate over a wide (0 V to 60 V) input common-mode voltage range. The internal amplifier can be used to limit the inrush current or to create a current source in a closed-loop system. The comparator can be used to monitor fault events for fast response. At power-up, the selectable op amp/comparator block is configured in the op amp mode. When the internal comparator is selected, the MAX9611 can be configured to have a latched-and-retry functionality, allowing a 60 V open-drain transistor output, ideal to operate high-side relay-disconnect FETs. Through the FDN360P MOSFET (element IC11 in Figure \ref{fig:SCH_Single_CH_FE}), the MAX9611 in comparator configuration can open the power line by limiting the current and turning off the SiPM. The current limit value is set through a DC voltage to the MAX9611. 

\begin{figure}[t]\center
\includegraphics[width=0.75\textwidth,keepaspectratio]{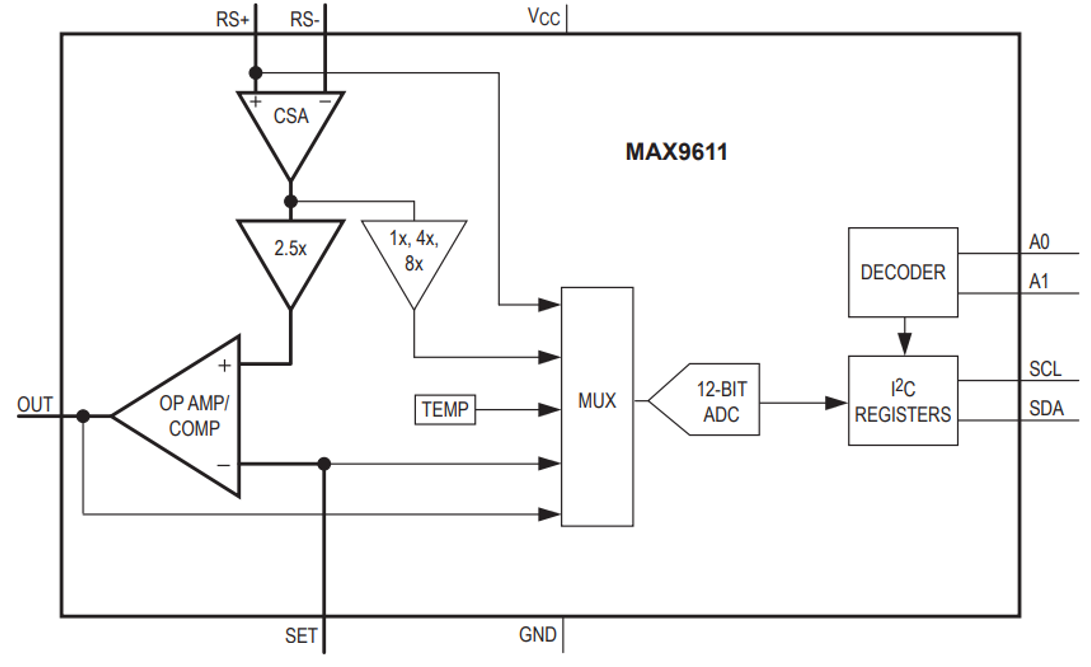}
\caption{Block diagram of the MAX9611 amplifier.}
\label{fig:BL_MAX9611_FE}
\end{figure}

To allow remote control of the current limit we added a partition network R131 and a variable resistance WR0 integrated in ISL23325TFVZ, which defines a reference voltage to the comparator. The comparator setting is calibrated between 0 and 3 $\mu$A of the SiPM bias current. An I$^{2}$C-controlled 12-bit, 500 sps ADC can be used to read the voltage across the sense resistor ($V_{SENSE}$), the input common-mode voltage ($V_{RSCM}$), the op amp/comparator output ($V_{OUT}$), the op amp/comparator reference voltage ($V_{SET}$), and an internal die temperature.  The ISL23325 is a low voltage, low noise, low power, dual DCP with 256 resistor step and an I$^{2}$C Bus interface. It integrates two DCP cores, wiper switches and control logic on a monolithic CMOS integrated circuit. The DCP can be used as a three-terminal potentiometer, the value of the resistance WR1 and WR2 change from 0 to 100 k$\Omega$. The block diagram of the MAX9611 is shown in Figure~\ref{fig:BL_MAX9611_FE}.

\subsection{Electronic Controller Board} \label{controller}

\begin{figure}[htbp!]\center
\includegraphics[width=1\textwidth,keepaspectratio]{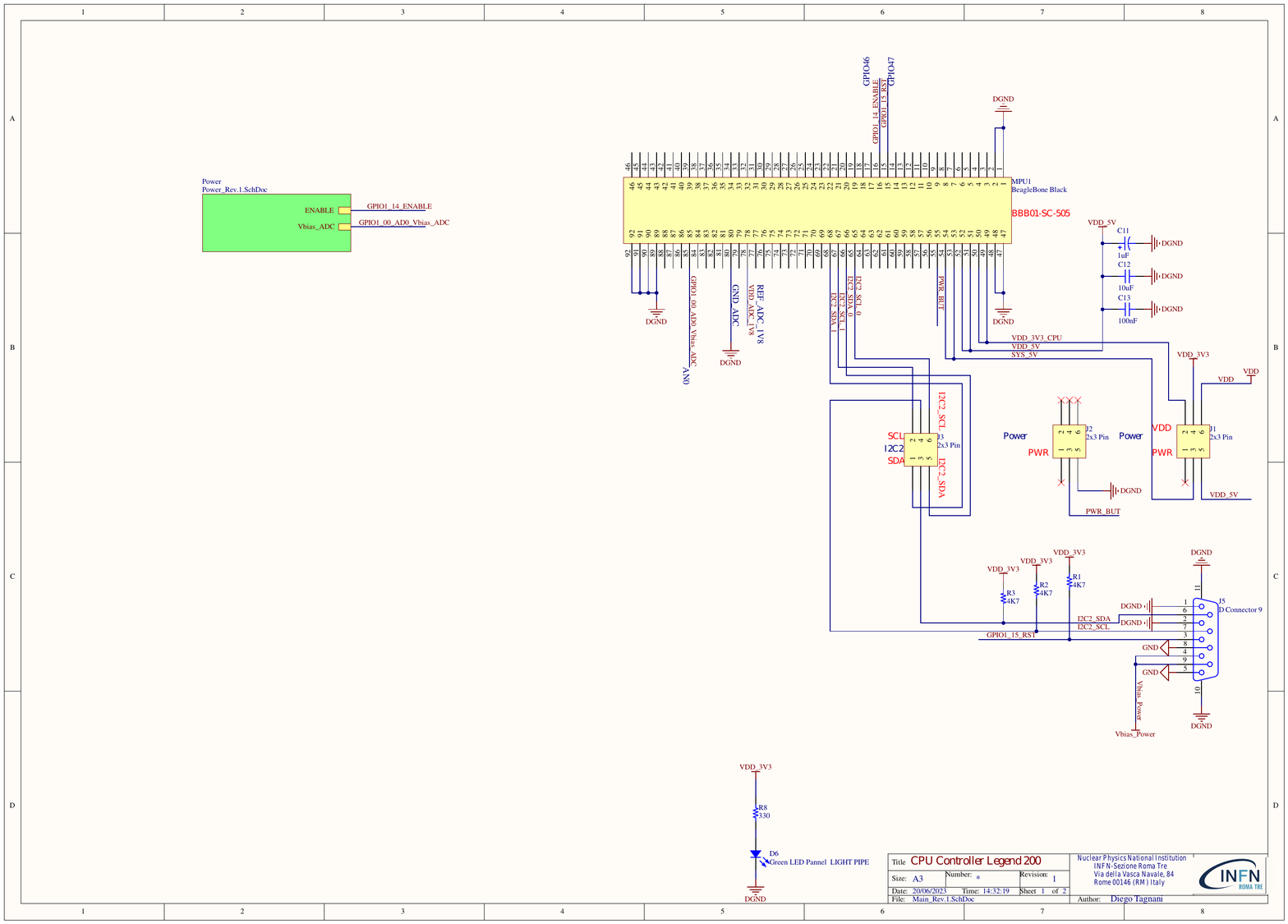}
\caption{Schematic main section of Controller board.}
\label{fig:SCH_Main_CO}
\end{figure}

A so called "controller board" is associated to the system. It contains the above mentioned intelligent Beaglebone Black MPU managing the entire slow control chain for all acquisition chains.
The BeagleBone Black is a product of the {\tt BeagleBoard.org} family.  
The Beaglebone Black MPU is placed inside the controller board with two multi-pin connectors and a mechanical support. The schematic main section of the controller board is displayed in Figure~\ref{fig:SCH_Main_CO}.

\begin{figure}[t]\center
\includegraphics[width=1\textwidth,keepaspectratio]{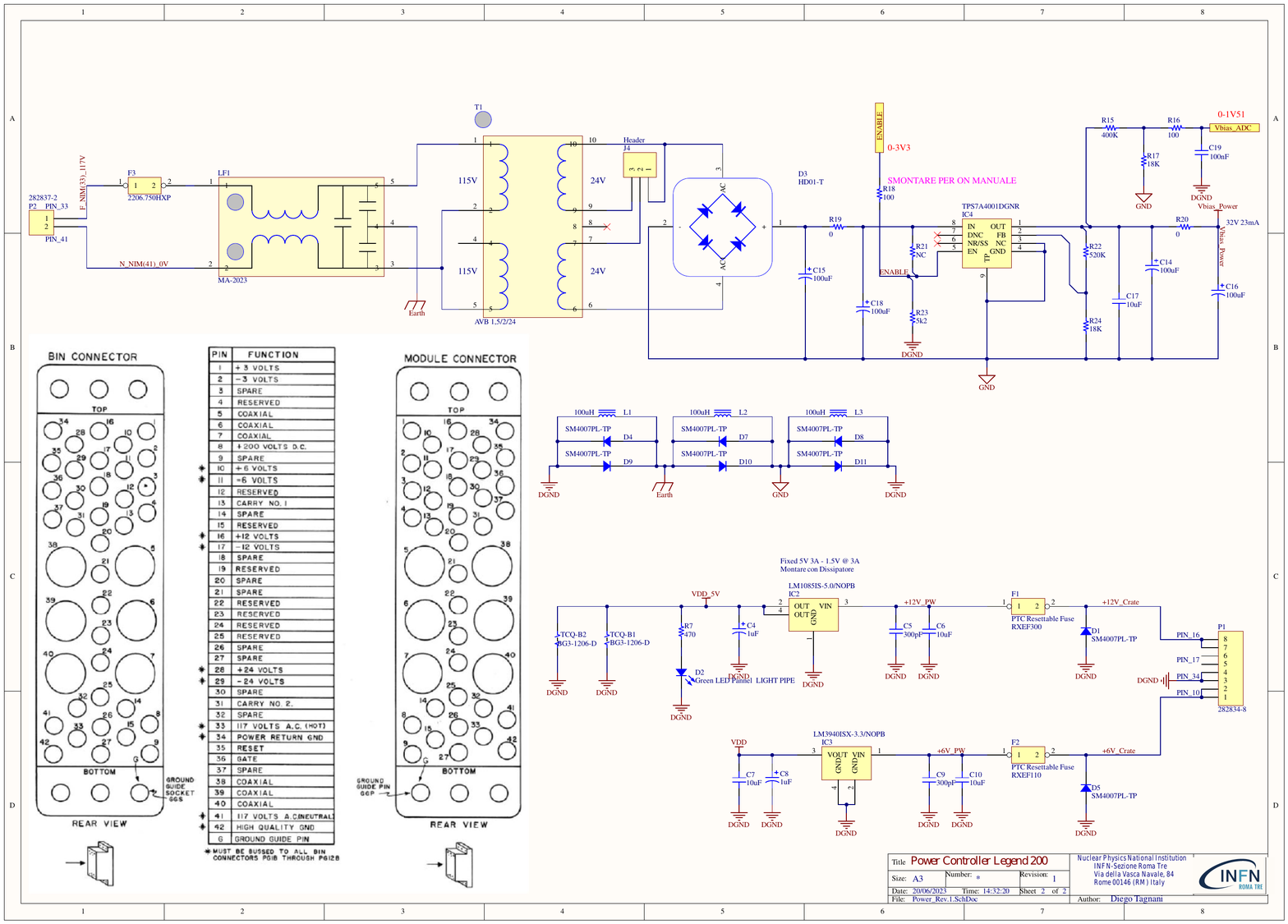}
\caption{Schematic power section of Controller board.}
\label{fig:SCH_Power_CO}
\end{figure}

\newpage
The controller board supplies power for the correct operation of the MPU. The voltages are regulated and filtered by DC +12 V and +6 V power supplies in the NIM standard, through the P1 connector. All the regulators are equipped with heat sinks and the crates are ventilated. This keeps the board operating temperature low, ensuring greater stability and reliability. We improve this further by adding BG3-1206-D thermal bridges. Each bridge connects the polygons of the board to its outline. It is exposed by the solder masks and put in contact with the standard NIM mechanics, to reduce its thermal resistance.
The I$^{2}$C, SDA, SCL and RESET bus wires are connected in a so-called daisy-chain through the J5 connector and a specific DB9 cable with 6 cascading connectors to the front-end cards. The same connector and cable also distribute the primary voltage, which is adjusted for each single channel of the FE cards. The primary voltage is generated by the control board through a AC-DC linear power supply. The schematic power section of the controller board is shown in Figure~\ref{fig:SCH_Power_CO}.

The NIM crate has a voltage of 117 $V_{AC}$ on the back-plane, it is protected by a fuse and filtered with a MA-2023 line filter. The board has an AVB1.5/2/24 transformer, which has two 115 $V_{AC}$ primary and two 24 $V_{AC}$ secondary. The jumper J4 allows to establish whether to use only a secondary or the series of the two. With the HD01-T rectifier bridge and the C15 and C18 buffer capacitors the voltage is rendered DC and then adjusted to the desired value through the linear voltage regulator TPS4001DGNR (schematic reference IC4). It has an enable input that is managed directly by the GPIO46 on the MPU. The R22 and R24 partition network allows to adjust the output voltage to 32 V$_{DC}$. This voltage is brought to the J5 connector on the internal plane of the PCB and has a guard line to minimize the noise. The same voltage is further attenuated by the network R15 and R17 and low-pass filtered by elements R16 and C19, near the connector of the MPU. Finally it is read with an internal ADC, it has a maximum dynamics of 1.8 V.

\section{ Slow control software for the LAr instrumentation }
This section describes the software architecture implemented to manage the electronic card and integrate the data with the LEGEND Central Slow Control (LCSC).

As described above, the electronics controller board is implemented with a BeagleBone Black device, where we have chosen to install an operating system based on the Ubuntu Linux distribution. A supplementary memory has been fitted, based on a 64GB SD card, to have enough available space to store programs, applications and data files.

Section \ref{II-A} describes the interaction and management of the electronic devices via I$^{2}$C and data management software architecture. Section \ref{II-B} describes the communication with LCSC via a TCP stream using a dedicated protocol. Section \ref{II-C} describes the calibration procedure for the slow control of the LAr instrumentation.

\subsection{Interaction and device management} \label{II-A}
The BeagleBone Black is practical as MPU since it allows to run Linux and the entire development stack directly on the control board, thus reducing latency and enhancing reliability. Python scripts can be developed to do thread management and communicate via serial buses (in particular I$^{2}$C). The software packages for I$^{2}$C communication were developed using Python 3.7.3 and the SMBus library \cite{ref:smbus}. Our software allows to enable and disable channels and set each channel voltage and current limit independently. It reads all parameters and the status of each channel. As previously described, the software can control 84 channels (five operating boards plus two spare boards) through the same I$^{2}$C bus. This is performed with routing strategies based on the management of I$^{2}$C multiplexers matching the experiment layout. Moreover, to reset the I$^{2}$C bus without rebooting the whole crate, we developed an emergency hard-reset routine that exploits a General Purpose Input/Output (GPIO).

Every channel of every board is addressed using a unique board identifier and a channel identifier. We set a floating-point value for the voltage and an integer value for each channel for the current limit (from 0 to 255). 

The control of the electronic devices runs at a rate greater than 10 Hz, compatible with the LEGEND requirement. For each iteration, the controller checks the values for every board and saves them in the data storage service, a Maria DB \cite{ref:maria} server. This data is then accessed to be sent to the LCSC, as described in the next section.

\subsection{Communication with LEGEND Central Slow Control} \label{II-B}

Parallel to the data acquisition and storage thread, another thread handles the communication with the LCSC. The latter looks for a trigger issued with a string, namely "SIPM:CONNECT", via TCP socket and collects data from the MariaDB instance to send it to the LCSC.

We developed this thread to be independent during data acquisition and to avoid interference with the other LEGEND readings and routing pipelines.

\subsection{Calibration of the LAr detector parameters} \label{II-C}

 Bus resistance differences combined with minor fabrication differences for two channels can result in significantly different output voltages recorded on the two channels, when triggered with the same hexadecimal value. For this reason, we calibrated every channel independently and generated a map between hexadecimal values and their measured output voltages. We changed the hexadecimal value on the ISL23325 while recording the change into a measured voltage using a multimeter controlled by a GPIB interface and with the MAX9611. We produced two different maps, one for setting the voltage (from voltage to hexadecimal value) and one for reading the recorded voltage (from hexadecimal value to voltage). Two distinct second-order polynomials were used to calibrate the response in both cases. The coefficients of the polynomials are stored in a database and sampled for every read/write instruction on the channels. 

\section{Overall noise level} \label{sec:noise-level}
After successfully integrating the front-end electronics with the experiment signal lines, we measured some key performance figures of the system. Results are as follows, with specific emphasis put on the electronic noise level. The electronic noise, which directly impacts the accuracy and reliability of the system, was measured systematically to provide quantitative insights into its performance. It was also measured in subsequent commissioning runs, and proved stable at the values showed below. 

\begin{figure}[b!]
\includegraphics[width=1.1\textwidth,keepaspectratio]{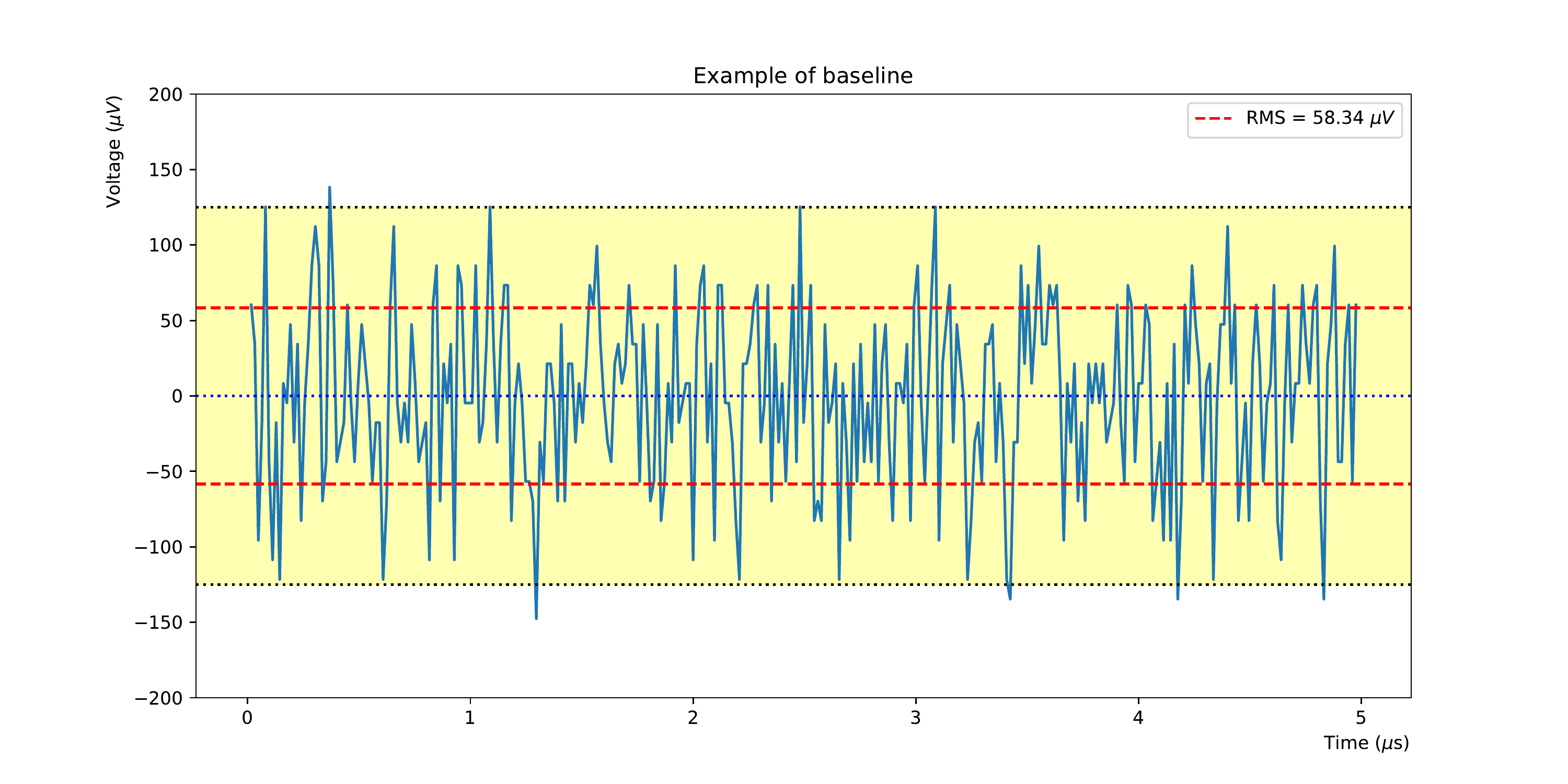}
\caption{Example of a baseline trace of 5 $\mu$s: the red horizontal lines represent the level of RMS for this specific trace (58.34 $\mu$V); while the yellow band, bounded by the two black dashed lines, represent the noise level of the baseline (250 $\mu$V, peak-to-peak).}
\label{fig:bls}
\end{figure}

Figure \ref{fig:bls} exemplifies a baseline from a SiPM connected to the front-end electronics, as described in the introduction of this work. The two red lines represent the Root Mean Square (RMS), while the yellow band indicates the overall noise level of the baseline, corresponding to 250 $\mu$V (peak-to-peak).

\begin{figure}[tp!]
\includegraphics[width=1.1\textwidth,keepaspectratio]{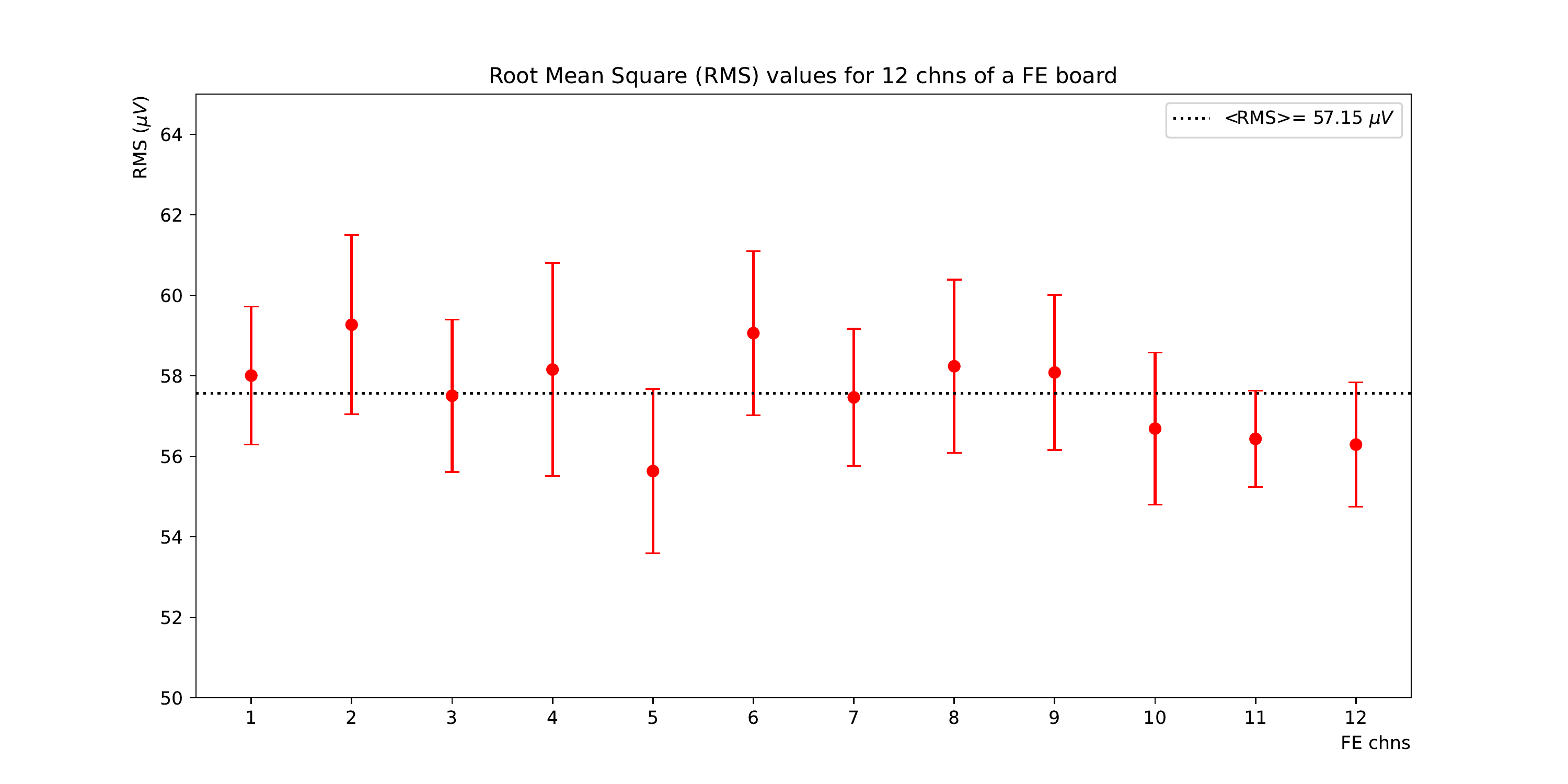}
\caption{RMS mean values of 3000 baselines for each 12 channels of a FE board. The black dashed line represents the RMS mean value between all channels.}
\label{fig:rms12}
\end{figure}
Additionally, Figure \ref{fig:rms12} illustrates the mean RMS level of 12 channels from the same front-end board. For each channel, we calculated the mean value using 3000 baselines. By determining the RMS value for each baseline and then averaging those values, we obtained the mean values reported on the plot. The black dashed line indicates the average RMS value across all channels. Notably, all channels exhibit a high level of agreement on this value, as depicted by the overlapping error bars.


\section{Conclusions}
This article provides a detailed description of the electronic front-end system for the liquid Argon instrumentation of the LEGEND-200 experiment at the Laboratori Nazionali del Gran Sasso. Based on the physics performance and technical requirements, an integrated system has been designed, produced and commissioned to receive and amplify the electrical signals from the silicon photo-multipliers immersed in the liquid Argon cryostat; and control their functional parameters. The system is characterized by a very low level of electrical noise, with an RMS of about 57 $\mu$V and a peak-to-peak excursion of around 250 $\mu$V. Such level was attained despite the many meters between the detectors and the front-end boards. These figures are compatible with the values in output from the circuit simulation and sport an excellent stability in time. The latter is also confirmed in the runs for physics, as part of the LEGEND-200 data taking chain.

\end{document}